\documentclass[preprint,superscriptaddress,showpacs,preprintnumbers,amsmath,amssymb]{revtex4}

\usepackage{booktabs}
\usepackage{graphicx,color}
\usepackage{dcolumn}
\usepackage{bm}
\usepackage{CJK}
\usepackage{mathrsfs}
\usepackage[dvipdfm,
            pdfstartview=FitH,
            CJKbookmarks=true,
            bookmarksnumbered=true,
            bookmarksopen=true,
            colorlinks,
            pdfborder=001,
            linkcolor=blue,
            anchorcolor=blue,
            citecolor=blue
            ]{hyperref}
\usepackage{doi}

\begin{document}
\begin{CJK}{GBK}{song}

\title{Spin symmetry in the Dirac sea derived from the bare nucleon-nucleon interaction}

\author{Shihang Shen}
\affiliation{State Key Laboratory of Nuclear Physics and Technology, School of Physics,
Peking University, Beijing 100871, China}
\affiliation{ Department of Physics, The University of Hong Kong, Pokfulam Road, Hong Kong, China}

\author{Haozhao Liang}
 \affiliation{RIKEN Nishina Center, Wako 351-0198, Japan}
 \affiliation{Department of Physics, Graduate School of Science, The University of Tokyo, Tokyo 113-0033, Japan}
\author{Jie Meng\footnote{Email: mengj@pku.edu.cn}}
 \affiliation{State Key Laboratory of Nuclear Physics and Technology, School of Physics,
Peking University, Beijing 100871, China}
 \affiliation{Department of Physics, University of Stellenbosch, Stellenbosch, South Africa}
 \affiliation{Yukawa Institute for Theoretical Physics, Kyoto University, Kyoto 606-8502, Japan}
\author{Peter Ring}
 \affiliation{State Key Laboratory of Nuclear Physics and Technology, School of Physics,
Peking University, Beijing 100871, China}
 \affiliation{Physik-Department der Technischen Universit\"at M\"unchen, D-85748 Garching, Germany}
\author{Shuangquan Zhang}
 \affiliation{State Key Laboratory of Nuclear Physics and Technology, School of Physics,
Peking University, Beijing 100871, China}

\date{\today}

\begin{abstract}
The spin symmetry in the Dirac sea has been investigated with relativistic Brueckner-Hartree-Fock theory using the bare nucleon-nucleon interaction.
Taking the nucleus $^{16}$O as an example and comparing the theoretical results with the data, the
definition of the single-particle potential in the Dirac sea is studied in detail.
It is found that if the single-particle states in the Dirac sea are treated as occupied states, the ground state properties are in better agreement with experimental data.
Moreover, in this case, the spin symmetry in the Dirac sea is better conserved and it is more consistent with the findings using phenomenological relativistic density functionals.
\end{abstract}


\maketitle


It is well known that in the nuclear system the spin symmetry is largely broken, that is, there exists a large spin-orbit (SO) splitting, which was introduced by Mayer~\cite{GoeppertMayer1949_PR75-1969} and Haxel et al.~\cite{Haxel1949_PR75-1766} in 1949. It formed the ground for the nuclear shell model. Twenty years later a new symmetry, the so-called pseudospin symmetry, was proposed to explain the near degeneracy between two single-particle (s.p.) states with the quantum numbers $(n,l,j=l+1/2)$ and $(n-1,l+2,j=l+3/2)$ \cite{Arima1969_PLB30-517,Hecht1969_NPA137-129}.
The two states are regarded as the pseudospin doublets with the pseudospin quantum numbers  $(\tilde{n} = n-1, \tilde{l}=l+1,j=\tilde{l}\pm 1/2)$.

By starting from the Dirac equation, it was found that the angular momentum of the pseudospin doublets $\tilde{l}$ is nothing but the orbital angular momentum of the lower component of the Dirac spinor, and the pseudospin symmetry is exact when the sum of vector and scalar potential $V+S$ vanishes \cite{Ginocchio1997_PRL78-436}.
The more general condition, $d(V+S)/dr = 0$, was proposed and can be approximately fulfilled in exotic nuclei \cite{Meng1998_PRC58-R628,Meng1999_PRC59-154}.
The general condition for spin and pseudospin symmetry, namely that $V+S$ is a constant for pseudospin symmetry is confirmed in Ref.~\cite{Ginocchio1999} and its connection to spin symmetry was also suggested there.
Since then, pseudospin symmetry has been realized as a relativistic symmetry and much work has been done to investigate its origin and its properties using phenomenological single-particle Hamiltonians, relativistic mean field theory, or relativistic Hartree-Fock (RHF) theory \cite{Marcos2001_PLB513-30,CHEN-TS2003_ChPL20-358,Lisboa2004_PRC69-024319,Long2006_PLB639-242, Long2010_PRC81-031302,Liang2011_PRC83-041301,Liang2013_PRC87-014334,SHEN-SH2013_PRC88-024311,
GUO-JY2014_PRL112-062502,SHI-M2014_PRC90-034318,ZHAO-Q2014_PRC90-054326,LI-DP2015_PRC91-024311, Tokmehdashi2015_ZNaturfA70-1,CHEN-S2016_SCG59-682011,Eshghi2016_EPJA52-201,LI-JJ2016_PRC93-054312, GAO-J2017_PLB769-77,SunTT2017}.

If one starts with a Dirac Hamiltonian, there exist single-particle states not only with positive energy but also with negative energy, states in the so-called Dirac sea.
It was shown in Ref.~\cite{ZHOU-SG2003_PRL91-262501} that the pseudospin symmetry in the positive spectrum has the same origin as the spin symmetry in the Dirac sea.
In other words, the SO doublets in the Dirac sea has the quantum number $(n,\tilde{l},j=\tilde{l}\pm 1/2)$, and the spin symmetry breaking term is proportional to $d(V+S)/dr$, similar to the pseudospin symmetry in the positive spectrum.
The spin symmetry in Dirac sea has also been investigated intensively afterwards \cite{HE-XT2006_EPJA28-265,SONG-CY2009_ChPL26-122102,Liang2010_EPJA44-119,Lisboa2010_PRC81-064324, Hamzavi2014_APNY341-153,SUN-M2017_IJMPE26-1750025}.
For comprehensive reviews on the study of pseudospin and spin symmetries, see Refs.~\cite{Ginocchio2005_PR414-165,Liang2015_PR570-1}.

Up until now, all the studies on the pseudospin symmetry in nuclei or the spin symmetry in the Dirac sea have been started from phenomenological s.p. Hamiltonians, or relativistic density functionals using phenomenological parameters~\cite{Ring1996_PPNP37-193,Vretenar2005_PR409-101,Meng2006_PPNP57-470, Niksic2011_PPNP66-519,Meng2016_IRNP10}.
It is therefore an interesting question to what extent spin symmetry in the Dirac sea is found in calculations starting from the bare nucleon-nucleon ($NN$) interaction which is fitted to the $NN$ scattering data and deuteron properties.
However, such \emph{ab initio} calculations for nuclei are extremely difficult and most of them are performed in a nonrelativistic framework~\cite{Dickhoff2004_PPNP52-377,LEE-D2009_PPNP63-117,LIU-Lang2012_PRC86-014302,
Barrett2013_PPNP69-131,Hagen2014_RPP77-096302,Carlson2015_RMP87-1067, Hergert2016_PR621-165}.
Only recently, a relativistic \emph{ab initio} method has been developed for finite nuclei by extending Brueckner-Hartree-Fock theory to the relativistic framework, and it has been shown that relativistic effects are important to improve the agreement with the experimental data~\cite{SHEN-SH2016_ChPL33-102103,SHEN-SH2017_PRC96-014316}.
In particular, the effect of tensor force is well treated in the spin-orbit splittings, as demonstrated in neutron drops~\cite{Shen2018}.

In this work, starting from a bare $NN$ interaction and taking the nucleus $^{16}$O as an example, we study the spin symmetry in the Dirac sea within relativistic Brueckner-Hartree-Fock (RBHF) theory.
Special attention will be paid on the definition of the s.p. potential in Dirac sea.
The results are compared with those obtained by phenomenological relativistic density functionals which are fitted to properties of finite nuclei and nuclear matter.


We use the relativistic version of the potential Bonn A. This is a relativistic one-boson-exchange $NN$ interaction which has been carefully adjusted to the $NN$ scattering data~\cite{Machleidt1989_ANP19-189}. The corresponding Hamiltonian has the form:
\begin{equation}
H=\sum_{kk^{\prime }}\langle k|T|k^{\prime }\rangle b_{k}^{\dagger
}b_{k^{\prime }}^{{}}+\frac{1}{2}\sum_{klk^{\prime }l^{\prime }}\langle
kl|V|k^{\prime }l^{\prime }\rangle b_{k}^{\dagger }b_{l}^{\dagger
}b_{l^{\prime }}^{{}}b_{k^{\prime }}^{{}},  \label{eq:hami}
\end{equation}%
where the relativistic matrix elements are given by
\begin{align}
\langle k|T|k^{\prime }\rangle & =\int d^{3}r\,\bar{\psi}_{k}(\mathbf{r})\left( -i%
\bm{\gamma}\cdot \nabla +M\right) \psi _{k^{\prime }}(\mathbf{r}), \\
\langle kl|V_{\alpha }|k^{\prime }l^{\prime }\rangle & =\int
d^{3}r_{1}d^{3}r_{2}\,\bar{\psi}_{k}(\mathbf{r}_{1})\Gamma _{\alpha }^{(1)}\psi
_{k^{\prime }}(\mathbf{r}_{1}) \label{eq:V} \notag \\
&~~~~~~~~\times D_{\alpha }(\mathbf{r}_{1},\mathbf{r}_{2})\bar{\psi}_{l}(\mathbf{r}_{2})\Gamma
_{\alpha }^{(2)}\psi _{l^{\prime }}(\mathbf{r}_{2}).
\end{align}
The indices $k,l$ run over a complete basis of Dirac spinors with positive and negative energies, as, for instance, over the eigensolutions of a Dirac equation with potentials of Woods-Saxon shape~\cite{ZHOU-SG2003_PRC68-034323,SHEN-SH2017_PRC96-014316,Meng2006}.

The two-body interaction $V_\alpha$ contains the exchange contributions of different mesons $\alpha= \sigma,\delta, \omega,\rho, \eta, \pi$. The interaction vertices $\Gamma _{\alpha }$ for particles 1 and 2 contain the corresponding $\gamma$-matrices for scalar $(\sigma,\delta)$, vector $(\omega,\rho)$, and pseudovector $(\eta, \pi)$ coupling and the isospin matrices $\vec\tau$ for the isovector mesons $\delta, \rho,$ and $\pi$.
For the Bonn interaction \cite{Machleidt1989_ANP19-189}, a form factor of monopole-type is attached to each vertex and $D_{\alpha }(\mathbf{r}_{1},\mathbf{r}_{2})$ represents the corresponding meson propagator. Retardation effects were deemed to be small and were ignored from the beginning.
Further details are found in Ref.~\cite{SHEN-SH2017_PRC96-014316}.

The matrix elements of the bare nucleon-nucleon interaction are very large and difficult to be used directly in nuclear many-body theory. Within Brueckner theory, the bare interaction is replaced by an effective interaction in the nuclear medium, the $G$-matrix. It takes into account the short-range correlations by summing up all the ladder diagrams of the bare interaction~\cite{Brueckner1954_PR95-217} and it is deduced from the Bethe-Goldstone equation~\cite{Bethe1957_PRSA238-551},
\begin{equation}
\bar{G}_{aba'b'}(W) = \bar{V}_{aba'b'}
 +\frac{1}{2}\sum_{cd} \frac{\bar{V}_{abcd}\bar{G}_{cda'b'}(W)}{W-e_{c}-e_{d}},
\label{eq:BG}
\end{equation}%
where $\bar{V}_{aba'b'}$ are the anti-symmetrized two-body matrix elements (\ref{eq:V}) and $W$ is the starting energy.
In self-consistent RBHF theory the states $|a\rangle, |b\rangle, ...$ are solutions of the relativistic Hartree-Fock (RHF) equations,
\begin{equation}
(T+U)|a\rangle =e_{a}|a\rangle ,
\label{eq:rhf}
\end{equation}%
where $e_{a}=\varepsilon _{a}+M$ is the s.p. energy with the rest mass of the nucleon $M$.
The intermediate states $c,\,d$ in Eq. (\ref{eq:BG}) run over all states above the Fermi surface with $e_c,\,e_d > e_F$, because the levels in the Fermi sea as well as those in the Dirac sea are occupied.

In the case of spherical symmetry, the s.p. wave function can be written as
\begin{equation} \label{eq:dirwf}
|a\rangle = \frac{1}{r} \left(%
\begin{array}{c}
F_{n_a\kappa_a}(r) \Omega_{j_am_a}^{l_a}(\theta,\varphi) \\
iG_{n_a\tilde{\kappa}_a}(r) \Omega_{j_am_a}^{\tilde{l}_a}(\theta,\varphi)%
\end{array}%
\right),
\end{equation}
where $\Omega_{jm}^l(\theta, \varphi)$ are the spinor spherical harmonics.
The radial, orbital angular momentum, total angular momentum, and magnetic
quantum numbers are denoted by $n,\, l,\, j,$ and $m$, respectively, while the
quantum number $\kappa$ is defined as $\kappa = \pm(j+1/2)$ for $j=l\mp1/2$.
Furthermore, $\tilde{l} = 2j - l$ is the orbital angular momentum for the lower component.
The corresponding effective local radial Dirac equation reads
\begin{equation}
\left(
\begin{array}{cc}
M+\Sigma(r) & -\frac{d}{dr}+\frac{\kappa}{r} \\
\frac{d}{dr}+\frac{\kappa}{r} & -M+\Delta(r)
\end{array}
\right) \left(
\begin{array}{c}
F_a(r) \\
G_a(r) \\
\end{array}
\right) =e_a \left(
\begin{array}{c}
F_a(r) \\
G_a(r) \\
\end{array}
\right),\label{eq:rdireq}
\end{equation}
with $\Sigma=V+S$ and $\Delta=V-S$ are the sum and difference of vector and scalar potentials.

The self-consistent s.p. potential $U$ in Eq.~(\ref{eq:rhf}) is defined by the $G$-matrix with the usual Hartree-Fock prescription. The problem is the starting energy $W$. Several methods have been introduced in the literature and we use here the method proposed in  Refs.~\cite{Baranger1969_Varenna40,Davies1969_PRC177-1519}. These were nonrelativistic investigations and therefore one had here only matrix elements $\langle a|U|b\rangle$ for s.p. states $|a\rangle$, $|b\rangle$ in the Fermi sea and above the Fermi level. In our earlier relativistic work~\cite{SHEN-SH2017_PRC96-014316} we treated in this context s.p. states $|a\rangle$, $|b\rangle$ in the Dirac sea as unoccupied, i.e. in a similar way as the states above the Fermi level. This leads to the following definition of the starting energy $W$ in the matrix elements of the self-consistent s.p. potential $U$:
\begin{widetext}
\begin{equation}\label{eq:u1}
\langle a|U|b\rangle=
\begin{cases}
\frac{1}{2}\sum_{i=1}^A \langle ai|\bar{G}(e_a+e_i)+ \bar{G}(e_b+e_i)|bi\rangle,
 & 0 < (e_a, e_b) \leq e_F \\
\sum_{i=1}^A \langle ai|\bar{G}(e_a+e_i)|bi\rangle,
& 0 < e_a \leq e_F,~e_b > e_F ~\text{or}~e_b < 0  \\
\sum_{i=1}^A \langle ai|\bar{G}(e'+e_i)|bi\rangle,
& e_a, e_b > e_F ~\text{or}~ < 0.
\end{cases}
\end{equation}
\end{widetext}
where the index $i$ runs over the occupied states in the Fermi sea (\emph{no-sea} approximation).
In the above equations, $e'$ is somewhat uncertain in the (R)BHF framework and it has been fixed as an energy among the occupied states in Ref.~\cite{SHEN-SH2017_PRC96-014316}.
The difference of the results by fixing $e'$ as the highest and as the lowest energy of the
occupied states in the Fermi sea has been discussed therein. As discussed in Ref.~\cite{SHEN-SH2017_PRC96-014316} the various matrix elements of the matrix $\bar{G}(W)$ are determined by interpolation and  with this choice the starting energy $W$ is limited as a sum of two single-particle energies in the Fermi sea.

From Eq.~(\ref{eq:u1}) it can be seen that in Ref.~\cite{SHEN-SH2017_PRC96-014316} the matrix elements $\langle a|U|b\rangle$ with s.p. states $|a\rangle$ and/or $|b\rangle$ in the Dirac sea (with $e < 0$) have been treated in the same way as those with states in unoccupied particle states ($e>e_F$). This is technically less time consuming as one does not need to calculate $\bar{G}(W)$ for values $W<0$. One should recall that there is no ``right'' or ``wrong'' choice for the s.p. potential in (R)BHF theory, as (R)BHF theory can be viewed as the 2 hole-line expansion in the more general hole-line expansion (or the Brueckner-Bethe-Goldstone expansion) \cite{Day1967_RMP39-719} and as the expansion goes to higher order the result becomes independent of the choice of $U$ \cite{SONG-HQ1998_PRL81-1584}.
On the other hand, there do exist ``better'' choices of $U$ as this choice will affect the convergence rate of the hole-line expansion.
It has been shown that the definition for hole states ($0 < e \leq e_F$) in Eq.~(\ref{eq:u1}) cancels a certain large amount of higher order diagrams thus it accelerates the convergence of hole-line expansion and improves the BHF approximation~\cite{Bethe1963_PR129-225,Baranger1969_Varenna40}, which corresponds to two hole lines.
However, there is no similar proof for the particle states nor for the states in the Dirac sea.
Thus, in the previous study of Ref.~\cite{SHEN-SH2017_PRC96-014316} they are chosen in a similar form as the hole states but with the uncertainty $e'$ in the starting energy in Eq.~(\ref{eq:u1}). This method will be labelled as ``previous'' in the following discussions.

In the present study, in the definition of the matrix elements $\langle a|U|b\rangle$, we will treat the s.p. states $|a\rangle$, $|b\rangle$ in the Dirac sea (with $e < 0$) as occupied (hole) states, which means the definition of the starting energy for the s.p. potential $U$ in Eq.~(\ref{eq:rhf}) becomes
\begin{widetext}
\begin{equation}\label{eq:u2}
\langle a|U|b\rangle=
\begin{cases}
\frac{1}{2}\sum_{i=1}^A \langle ai|\bar{G}(e_a+e_i)+ \bar{G}(e_b+e_i)|bi\rangle,
 & e_a, e_b \leq e_F \\
\sum_{i=1}^A \langle ai|\bar{G}(e_a+e_i)|bi\rangle,
& e_a \leq e_F,~e_b > e_F  \\
\sum_{i=1}^A \langle ai|\bar{G}(e'+e_i)|bi\rangle,
& e_a, e_b > e_F.
\end{cases}
\end{equation}
\end{widetext}
This choice seems to be reasonable since in the Bethe-Goldstone equation (\ref{eq:BG}) the intermediate states $c,d$ are only allowed to be states above the Fermi surface $e_c,\,e_d>e_F$.
From this point of view, the s.p. states in the Dirac sea are ``occupied'' hole states.

The calculation based on Eq.~(\ref{eq:u2}) will be labelled as ``present''.
In the following discussions we will compare the results of RBHF calculations using the previous definition~\cite{SHEN-SH2017_PRC96-014316} of the s.p. potential $U$ in Eq. (\ref{eq:u1})  with those using the present defintion in Eq. (\ref{eq:u2}).
As the difference between these two definitions affects mainly the states in the Dirac sea, we expect changes mostly for the s.p. properties in the Dirac sea.
The Bonn A interaction \cite{Machleidt1989_ANP19-189} will be used, and the nucleus $^{16}$O is taken as an example. All the other numerical details are the same as in the previous study of Ref.~\cite{SHEN-SH2017_PRC96-014316}. We use in all cases $e' = e_{\pi 1p1/2}$.

\begin{table}[!th]
\caption{Total energy $E$, rms charge radius $r_c$, and proton $1p$ spin-orbit splitting $\Delta E_{\pi 1p}^{ls}$ of $^{16}$O.  Results of RBHF calculations with different definitions of the starting energy in the definition of the potential $U$ are compared with the data~\cite{Wang2017_ChPC41-030003,Angeli2013_ADNDT99-69,Coraggio2003_PRC68-034320}.}
\label{table1}
\centering
\begin{ruledtabular}
\begin{tabular}{lccc}
& Previous~\cite{SHEN-SH2017_PRC96-014316} & Present & Exp. \\
\hline
$E$ (MeV) & $-113.5$ & $-120.2$ & $-127.6$ \\
$r_c$ (fm) & $2.56$ & $2.53$ & $2.70$ \\
$\Delta E_{\pi 1p}^{ls}$ (MeV) & $5.4$ & $5.3$ & $6.3$ \\
\end{tabular}
\end{ruledtabular}
\end{table}

In Table \ref{table1} we show the total energy, the rms charge radius, and the proton $1p$ spin-orbit splitting of $^{16}$O. RBHF calculations with the interaction Bonn A and two choices for the starting energy in the potential $U$ are compared with experimental data~\cite{Wang2017_ChPC41-030003,Angeli2013_ADNDT99-69,Coraggio2003_PRC68-034320}:
(I) previous definition~\cite{SHEN-SH2017_PRC96-014316} in Eq.~(\ref{eq:u1}) and (II)  present definition in Eq.~(\ref{eq:u2}).
The present total energy $E = -120.2$ MeV gives nearly $7$ MeV more binding than the previous result and is in better agreement with the data. On the other hand, the rms charge radius is by $0.03$ fm smaller than the previous result, and the SO splitting is smaller by $0.1$ MeV.

\begin{figure}
\includegraphics[width=8cm]{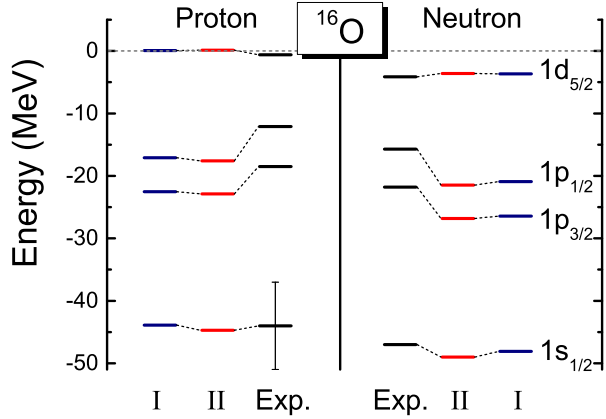}
\caption{(Color online) Single-particle spectrum in the Fermi sea of $^{16}$O calculated by RBHF with different choices of s.p. potential in the Dirac sea (I for previous~\cite{SHEN-SH2017_PRC96-014316} and II for present choice), in comparison with experimental data \cite{Coraggio2003_PRC68-034320}.}
\label{fig1}
\end{figure}

Fig.~\ref{fig1} shows the s.p. spectrum in the Fermi sea of $^{16}$O calculated by RBHF theory with different choices of the s.p. potential $U$ in the Dirac sea, in comparison with experimental data \cite{Coraggio2003_PRC68-034320}.
With the present choice of $U$, the s.p. energies are lower than the previous results. This leads to a more bound and smaller nucleus as shown in Table \ref{table1}.
As has already been discussed in Ref.~\cite{SHEN-SH2016_ChPL33-102103}, the $p$ levels are slightly too low as compared with the data.
This might be due to the lack of more complicated configurations such as particle vibration coupling~\cite{VinhMau1995_NPA592-33,Litvinova2006_PRC73-044328} in the RBHF framework, where only the ladder diagrams have been included.

\begin{figure}
\includegraphics[width=6.5cm]{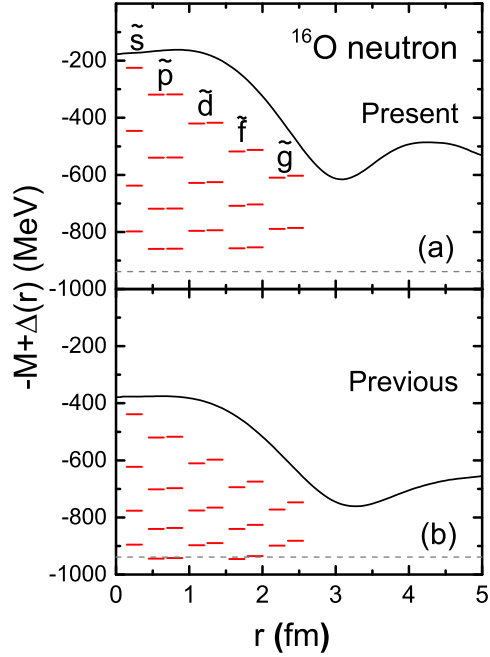}
\caption{(Color online) Single-particle spectrum in the Dirac sea and the effective single-particle potential for the $\nu 1s1/2$ channel in the Dirac sea calculated by RBHF theory using the interaction Bonn A  with (a) the present and (b) the previous~\cite{SHEN-SH2017_PRC96-014316} choice of s.p. potential in the Dirac sea.}
\label{fig2}
\end{figure}

In Fig.~\ref{fig2}, we show the s.p. spectrum and the effective single-particle potential $\Delta(r) = V(r) - S(r)$ in the Dirac sea calculated by RBHF theory with different choices of the starting energy in the s.p. potential $U$. The s.p. levels are grouped by the angular momentum $\tilde{l}$ of the lower component in the Dirac spinor (\ref{eq:dirwf}) with negative energy, thus, $\tilde{s}, \tilde{p}, \tilde{d}, \dots$ means $\tilde{l} = 0, 1, 2, \dots$. We consider in the following the spin-orbit (SO) splitting of these levels.
The potentials in both panels are not approaching 0 when $r\to \infty$ as usually found in the RMF study \cite{ZHOU-SG2003_PRL91-262501} because of the nonlocality of the RBHF s.p. potential $U$ in Eq.~(\ref{eq:rhf}).
Different s.p. wave function will give different effective s.p. potentials, and the one shown in Fig.~\ref{fig2} is calculated from the wave function of $\nu 1s1/2$ (or $\nu 1\tilde{p}1/2$ if labelled with the angular momentum of lower component in Eq.~(\ref{eq:dirwf})) using
\begin{equation}
\Delta(r) = e_a + M - \frac{\frac{dF_a(r)}{dr}+\frac{\kappa}{r}F_a(r)}{G_a(r)},
\end{equation}
which can be derived from the effective local radial Dirac equation (\ref{eq:rdireq}).

By comparing panel (a) and (b) in Fig.~\ref{fig2}, it can be seen that present calculation gives a deeper s.p. potential in the Dirac sea, and the spectra are higher by $100\sim 200$ MeV.
Moreover, the SO splittings in the present results are generally smaller.

\begin{figure}[!thbp]
\includegraphics[width=8cm]{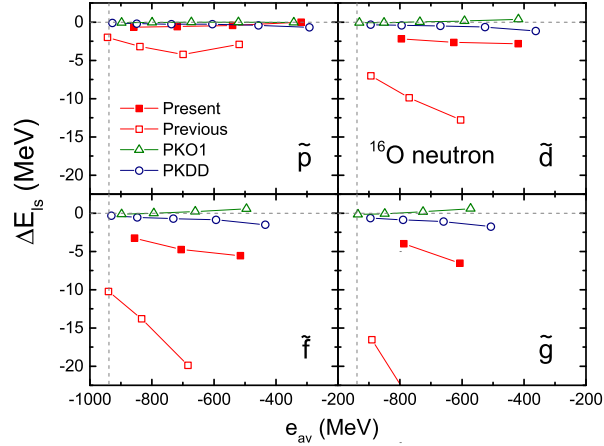}
\caption{(Color online) SO splittings $\Delta E_{\rm ls} = e_{j_<} - e_{j_>}$ versus the average energy of the SO doublets calculated by RBHF with previous~\cite{SHEN-SH2017_PRC96-014316} and present choices of single-particle potential in the Dirac sea, in comparison with results of relativistic density functionals PKDD \cite{Long2004_PRC69-034319} and PKO1 \cite{Long2006_PLB640-150}.}
\label{fig3}
\end{figure}

In order to see the SO splittings more clearly, we show in Fig.~\ref{fig3} the SO splittings $\Delta E_{\rm ls} = e_{j_<} - e_{j_>}$ versus the average energy of the SO doublets $e_{\rm av} = (e_{j_<} + e_{j_>})/2$, where $j_< = \tilde{l} - 1/2$ and $j_> = \tilde{l} + 1/2$.
The results are compared with those of phenomenological relativistic density functionals PKDD \cite{Long2004_PRC69-034319} and PKO1 \cite{Long2006_PLB640-150}.
With present choice, the SO splittings calculated by RBHF are much smaller thus the spin symmetry is better conserved, which is in better agreement with phenomenological relativistic density functional findings.
However, for SO doublets with large angular momentum such as $\tilde{f}$ and $\tilde{g}$, the SO splittings given by RBHF are still quite large comparing with PKDD or PKO1.

\begin{figure}[b]
\includegraphics[width=8cm]{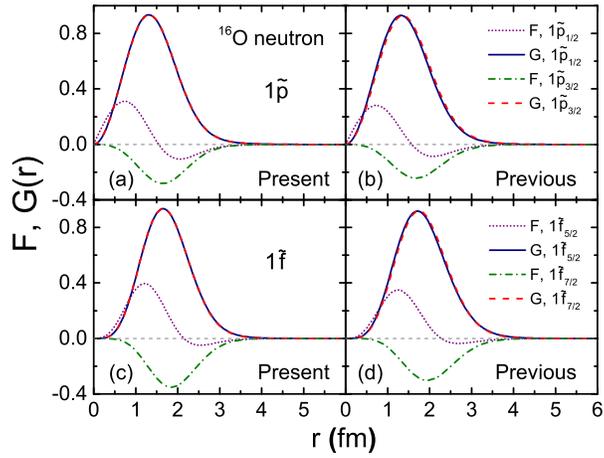}
\caption{(Color online) Wave functions of SO doublets $1\tilde{p}$ and $1\tilde{f}$ calculated by RBHF with the present and the previous~\cite{SHEN-SH2017_PRC96-014316} choice of single-particle potential in the Dirac sea.}
\label{fig4}
\end{figure}

As an example, we show in Fig.~\ref{fig4} the wave functions of the SO doublets $1\tilde{p}$ and $1\tilde{f}$ calculated by RBHF with the present and the previous choices.
Unlike for the states with positive energy, the upper component $F(r)$ of states with negative energy is the small component and the lower component $G(r)$ is the large component.
For a given SO doublet in the Dirac sea such as $1\tilde{p}$, the lower components are very close to each other as the SO splitting can be treated as a small perturbation.
Correspondingly, when the SO splitting increases, which is the case for the previous choice of $U$~\cite{SHEN-SH2017_PRC96-014316}, the difference of $G(r)$ between the SO doublets also increases as shown in panels (b) and (d).

In summary, we have studied the spin symmetry in the Dirac sea with the bare $NN$ interaction Bonn A using relativistic Brueckner-Hartree-Fock theory. No three-body forces have been taken into account. Different choices of the starting energy in the single-particle potential of the Dirac sea have been investigated.
It has been found that, if the single-particle states in the Dirac sea are treated as occupied hole states, the ground state energy of $^{16}$O calculated by RBHF theory is in better agreement with experimental data, while the charge radius and spin-orbit splittings are slightly worse than in the earlier calculations~\cite{SHEN-SH2017_PRC96-014316}, where they have been treated as empty states.
Furthermore, the spin symmetry is much better conserved with this choice. This is also more consistent with findings of phenomenological relativistic density functional theory.
Therefore, it is suggested to use this definition of the single-particle potential in the Dirac sea in future RBHF investigations.
In the present results, the SO splittings with higher angular momentum are still quite large compared with those obtained with phenomenological relativistic density functionals.
One may try to investigate in detail how different channels of the effective interaction $G$-matrix contribute to the spin symmetry in the Dirac sea, such as the scalar, vector, and tensor channels.
In the future, it is also interesting to see how different bare interactions will influence the results, such as a relativistic chiral interaction \cite{REN-XL2017_CPC,LI-KW2016}.

\section*{ACKNOWLEDGMENTS}

This work was partly supported by the Major State 973 Program of China No.~2013CB834400, Natural Science Foundation of China under Grants No.~11335002, No.~11375015, and No.~11621131001,  the Overseas Distinguished Professor Project from Ministry of Education of China No.
MS2010BJDX001, the Research Fund for the Doctoral Program of Higher Education of China under Grant No.~20110001110087, and the DFG (Germany) cluster of excellence \textquotedblleft Origin and Structure of the Universe\textquotedblright\ (www.universe-cluster.de).
H.L. would like to thank the RIKEN iTHES project and iTHEMS program.

\bibliographystyle{apsrev4-1}

\end{CJK}
\end{document}